\documentstyle[pra,aps,epsfig,amssymb,multicol]{revtex}

\begin{document}

%\draft

\title{Spin Squeezing in the Ising Model}

\author{Xiaoguang Wang, Anders S\o ndberg S\o rensen and Klaus M\o lmer}

\address{Institute of Physics and Astronomy, University of Aarhus , \\
DK-8000, Aarhus C, Denmark}

\date{\today}

\maketitle

\begin{abstract}

We analyze the collective spin noise in 
interacting spin systems. 
General expressions are derived for the short time 
behaviour of spin systems with general spin-spin interactions, 
and we  suggest optimum
experimental conditions for the detection of spin squeezing.
For Ising models with site dependent nearest neighbour 
interactions general expressions are presented
for the spin squeezing parameter for all times.
The reduction of collective spin noise can be used to
verify the entangling powers of quantum computer architectures 
based on interacting spins. 

\end{abstract}


\section{Introduction}

Parallel with the study of non-classical squeezed states of electromagnetic
radiation \cite{Squeezed}, increasing attention has been devoted 
to the study of atomic spin squeezed states 
\cite{Kitagawa,Wineland,Agarwal,Kuzmich1,Lukin,Vernac,Kuzmichqnd,%
Soerensen1,Soerensen3,Soerensen2,hald,Kuzmichexp,kasevich}.
In Ramsey spectroscopy a sample of $N$ two-level atoms is represented by the
collective operators $J_\alpha =\sum_{i=1}^Nj_{\alpha ,i}$ ($\alpha=x$,
$y$, or $z$), where $j_{\alpha
,i}=\sigma _{\alpha ,i}/2$ and $\sigma _{\alpha ,i}$ are the Pauli operators
for the $i^{th}$ atom. 
Atomic spin squeezed states are quantum correlated states with 
reduced fluctuations in one of the collective spin components, and they
have possible applications in  
atomic interferometers and high precision atomic clocks.
Wineland et al. \cite{Wineland} have shown that the frequency resolution in
spectroscopy depends on the spin squeezing parameter
\begin{equation}
\xi ^2=\frac{N\left( \Delta J_\perp \right) ^2}{\langle
  \vec{J}\hspace{0.1cm}\rangle ^2} ,
\label{eq:squeeze}
\end{equation}
where $\perp$ denotes a direction perpendicular to the mean spin. The
inequality $\xi ^2<1$ 
indicates that the system is spin squeezed, and it has been shown that
any state with $\xi^2<1$ is an entangled state
\cite{Soerensen3,Soerensen2}.

In 1993 Kitagawa and Ueda showed that spin squeezing is produced by simple
nonlinear 
spin Hamiltonians in analogy with the non-linear 
Hamiltonians leading to squeezed light \cite{Kitagawa}.
A number of experimental proposals for atomic spin squeezing
have appeared involving interaction of atoms with squeezed 
light \cite{Wineland,Agarwal,Kuzmich1,Lukin,Vernac}, 
quantum non-demolition measurement of atomic spin states\cite{Kuzmichqnd}, 
and atomic collisional interactions \cite{Soerensen1,Soerensen3},
and recently the first experimental realizations of spin squeezing have
been achieved 
\cite{hald,Kuzmichexp,kasevich}. 

There is a  link
between the theory of quantum computing and the theory of spin
squeezing through the observation that the register of
quantum bits in a quantum computer constitutes an ensemble of
effective spins, and a general
purpose quantum computer can  obviously produce a spin squeezed state
of its register qubits. Moreover, proposals for quantum computing
may be partly implemented, e.g., with only a restricted class of
operations, which do not suffice for general computing purposes,
but which do lead to massive entanglement and possibly to
spin squeezing. 
We have previously considered such
{\it reduced instruction set} (RISQ) quantum computers with ions
and atoms \cite{risq}, 
and we have pointed out that apart from their ability to
address particular physics problem such as 
anti-ferromagnetism, they can also be used to synthesize useful
quantum states. 
In this paper we present a theory for the spin squeezing
expected for different models of interacting spins. These
models encompass a number of theoretical proposals for quantum
computing, and although spin squeezing may not be
a particularly relevant property in, e.g., a spintronic or a 
quantum dot realization of a quantum computer, we wish to point out 
the possibility of verifying the entangling powers of the 
permanent or controllable interactions in these systems by
simple measurements on the entire system. The detection of spin squeezing
may constitute a useful diagnostic tool in the early stages of the
construction of a quantum computer.

The treatment of the most general case of interacting spin
systems is a formidable task, dealt with by a number of ingenious
approximations in the theory of magnetism in solid state physics.
In section \ref{general} of the paper, we consider the case of an initially
known state of the spins, subject to an arbitrary interaction
Hamiltonian. By application of Ehrenfest's theorem, we can determine
the short time behaviour of the system analytically, we can identify the 
spin squeezing signal, and we can devise the optimum conditions
for detecting this signal. In Section \ref{ising}, we consider a general Ising
type Hamiltonian with the spins constituting a chain with only nearest
neighbour interactions of varying magnitude. We obtain an analytical
expression for the spin squeezing of the system at any future time,
and we provide different examples of spin chains with constant,
alternating and random couplings.

\section{Short time behaviour for general pairwise interactions} 
\label{general}
Consider a collection of spin 1/2 particles which only
interacts through pairwise interactions.  
The most general Hamiltonian describing this situation is given by
\begin{equation}
H=\sum_{k\neq l} \vec{j}_k \hspace{0.0cm}^{\mathbb T} \cdot {\underline {\underline
    m}}^{kl} 
\cdot \vec{j}_l, 
\label{generalham}
\end{equation}
where $^{\mathbb T}$ denotes the transpose $\vec{j}_k \hspace{0.1cm}^{\mathbb
  T}=(j_{x,k},j_{y,k},j_{z,k})$, and where
${\underline {\underline  m}}^{kl}$ are real 3 by 3 matrices.
We assume that the particles  are initially 
prepared 
in a state where all spins are pointing in a given direction 
prepared for instance by optical pumping techniques, and 
we compute the time
evolution of the noise in a component perpendicular to the
direction of the mean spin. 
A convenient representation of this situation
is by a collective spin state 
$\hat{R}(\alpha,\beta,\gamma)|JJ\rangle$, where $\hat{R}$ is a rotation
operator given by the  three Euler angles $\alpha$, $\beta$, and
$\gamma$, and where $|JM\rangle$ denotes an eigenstate of
  $\vec{J}\hspace{0.1cm}^2$  and
$J_z$ with eigenvalues $J(J+1)$ and $M$. The noise in a direction
perpendicular to the mean spin can be found by calculating the square of
the rotated $J_x$ operator $J_\perp=
\hat{R} J_x \hat{R}^\dagger$.    

With the general Hamiltonian (\ref{generalham}) it is not possible to
calculate the full time evolution of the noise. It is however straightforward
by Ehrenfest's theorem to determine the short time evolution 
\begin{eqnarray}
\frac{d}{dt} (\Delta J_\perp)^2 
%&=&\frac{i}{\hbar}\langle
%JJ|\hat{R}^\dagger [H,\hat{R} J_x^2 \hat{R}^\dagger] \hat{R}|JJ\rangle
%\nonumber \\
=\frac{i}{\hbar}\langle
JJ| [\tilde{H}, J_x^2]|JJ\rangle,
\label{ehrenfest}
\end{eqnarray} 
where we have introduced the transformed Hamiltonian 
$\tilde{H}=\hat{R}^\dagger
H\hat{R}$ and we have used $\langle J_\perp \rangle$=0. By using
$\hat{R}^\dagger  
\vec{J}\hat{R} = {\underline{\underline R}}\cdot \vec{J}$, where
${\underline {\underline R}}$ is the 3 by 3 matrix representation of the
rotation in coordinate space, we may express the transformed Hamiltonian as 
$\tilde{H}=\sum_{k\neq l} \vec{j}_k \hspace{0.0cm}^{\mathbb T} \cdot \tilde{{\underline
  {\underline  m}}}^{kl} \cdot \vec{j}_l$, where $\tilde{{\underline
  {\underline  m}}}^{kl}={\underline {\underline R}}^{\mathbb T}
\cdot 
{\underline {\underline  m}}^{kl}\cdot {\underline {\underline R}}$. By
calculating the commutator and evaluating the expectation value we find
\begin{equation}
\frac{d}{dt} (\Delta J_\perp)^2 = \frac{1}{2\hbar} (0,1,0)\cdot
\tilde{ {\underline {\underline M}}}\cdot {\left( \begin{array}[c]{c} 1 \\
      0 \\ 0  
\end{array} \right)}.
\label{derivative}
\end{equation}
In this expression we have introduced a matrix ${\underline {\underline M}}
= \sum_{k\neq l} {\underline {\underline m}}^{k,l}$, and transformed it
with the rotation matrix $\tilde{ {\underline {\underline M}}}={\underline
  {\underline R}}^{\mathbb T} \cdot
 {\underline {\underline M}}\cdot {\underline {\underline R}}$. 

In Eq.~(\ref{generalham}) the terms involving the $k$th and $l$th
particles are  $\vec{j}_k \hspace{0.0cm}^{\mathbb T} \cdot {\underline
  {\underline  m}}^{kl} \cdot \vec{j}_l$ and $ \vec{j}_l
\hspace{0.0cm}^{\mathbb T} \cdot {\underline {\underline 
    m}}^{lk} 
\cdot \vec{j}_k$. The sum of these terms can also be written as
$1/2(\vec{j}_k \hspace{0.0cm}^{\mathbb T} \cdot ({\underline 
  {\underline  m}}^{kl}+{\underline 
  {\underline  m}}^{lk \ {\mathbb T}}) \cdot \vec{j}_l + \vec{j}_l
\hspace{0.0cm}^{\mathbb T} \cdot ( {\underline 
  {\underline  m}}^{lk}+{\underline 
  {\underline  m}}^{kl\ {\mathbb T}})
\cdot \vec{j}_k)$. From this expression we see that without loss of
generality we can assume the matrix ${\underline {\underline M}}$ to be
symmetric and hence diagonalizable and in a convenient coordinate
system we have 
\begin{equation}
{\underline {\underline M}}={\left(
\begin{array}[c]{ccc}
M_x & 0 & 0\\
0&M_y&0\\
0&0&M_z
\end{array}
\right)}.
\label{m}
\end{equation}

By calculating $\tilde{{\underline {\underline M}}}$ using the form
(\ref{m}) for the matrix ${\underline {\underline M}}$ and the well known
rotation matrices ${\underline {\underline R}}$, we can find the time
derivative of the noise perpendicular to the mean spin for any orientation
of the spin by using Eq.~(\ref{derivative}). Since we start out in a state
with $\xi^2=1$ and since $d/dt \langle \vec{J}\hspace{0.1cm}\rangle^2=0$ at
$t=0$,  the interaction produces spin squeezing
if we can find any set of 
Euler angles $\alpha$, $\beta$, and $\gamma$ which gives a negative
derivative in Eq. (\ref{derivative}).  The optimal orientation of the spin
is found by minimizing the derivative (\ref{derivative}) with respect
to the angles  $\alpha$, $\beta$, and $\gamma$. 
We find that the extrema of Eq.~(\ref{derivative}) are always with the 
mean spin along  one of the eigenvectors of ${\underline {\underline
M}}$.  If the
spin is polarized along the $z$-axis the change in the noise is maximal for
the perpendicular components $J_{\pm\pi/4}=1/\sqrt{2}(J_x\pm J_y)$ and we
find 
\begin{equation}
\frac{d}{dt} (\Delta J_{\pm \pi/4})^2 =\pm \frac{1}{4\hbar}(M_y-M_x).
\label{ddtpi/4}
\end{equation}
Similar expressions are found if the spin is oriented along the $x$ or
$y$-axes. 

Note that no assumptions about the values of the coupling matrices are
made, they may vary randomly for any pair $(k,l)$ of spins. Only the sum
needs to be specified to determine the short time spin squeezing.
The above argument only states that some squeezing will be
produced, it does not say anything about the maximum
squeezing. In the following section we shall analyze special cases
of the interaction (\ref{generalham}) where the squeezing can be calculated
exactly for all times.  

\section{Ising chain with nearest neighbour coupling}
\label{ising}

Consider now a general model with arbitrary coupling constants
between nearest neighbours in an Ising spin chain. The chain
consists of $N$ spins with the Hamiltonian
\begin{equation}
H=\hbar \sum_{i=1}^{N} \chi _ij_{x,i}j_{x,i+1},
\label{hamil}
\end{equation}
where  we identify the $N+1^{st}$ spin with the first one in the
chain. 
Depending on the value of $\chi_N$, the chain can be equipped with
open or closed boundary conditions.

This Hamiltonian arises in recent proposals for quantum
computation with atoms in optical lattices 
\cite{innsbruck_lattice,jessen}. In these proposals
the atoms interacts with the nearest neighbours and it has been shown that
the 
interaction can be put in the form (\ref{hamil}) and that this interaction
produces spin squeezing \cite{Soerensen1}. Other application of this
Hamiltonian in optical lattices can be found in \cite{briegel}.
Spin squeezing  is
much less experimentally challenging to produce and to verify than 
the full quantum computer, and
squeezing can provide a demonstration of the entangling capabilities of
the setup as well as having practical application in atomic clocks. From
the discussion in section \ref{general} follows that
for short times the optimal squeezing is produced by having the spin
initially
polarized along $z$ and by looking at the noise in one of the 
components  $J_{\pm\pi/4}=1/\sqrt{2}(J_x\pm J_y)$, where
$\frac{d}{dt} (\Delta J_{\pm \pi/4})^2 =\mp \frac{1}{4}\sum_i \chi_i$.
For longer time intervals the optimal noise reduction is in a component
$J_\theta= 
\cos (\theta) J_x+\sin (\theta)J_y$ with $\theta\neq \pm \pi /4$ but for
simplicity we shall only consider $\theta=\pm \pi /4$.

The unitary time evolution operator for the Hamiltonian $H$ can be
written

\begin{equation}
U(t)=e^{-iHt/\hbar}=
\prod_{i=1}^{N}e^{-i\chi_i t j _{x,i} j _{x,i+1}},
\label{eq:u}
\end{equation}
and the operators $\vec{j}_{i}(t)$ in the Heisenberg picture are
readily obtained

\begin{equation}
\left( \begin{array}{c} j_{x,i}(t)\\j_{y,i}(t)\\j_{z,i}(t) 
\end{array} \right)
= \left( \begin{array}{ccc} 1 & 0 & 0 \\
0 & \cos \left( \chi _{i-1}tj_{x,i-1}+\chi _itj_{x,i+1}\right) & -\sin
\left(
\chi _{i-1}tj_{x,i-1}+\chi _itj_{x,i+1}\right) \\
0 & \sin \left( \chi _{i-1}tj_{x,i-1}+\chi _itj_{x,i+1}\right) & \cos
\left(
\chi _{i-1}tj_{x,i-1}+\chi _itj_{x,i+1}\right) \end{array} \right)
\cdot
\left( \begin{array}{c} j_{x,i}(0)\\j_{y,i}(0)\\j_{z,i}(0) 
\end{array} \right).
%TCIMACRO{\binom{j_{y,i}(t)}{j_{z,i}(t)} }
%BeginExpansion
%{j_{y,i}(t) \choose j_{z,i}(t)}%
%%EndExpansion
%=\left( 
%\begin{array}{ll}
%\cos \left( \chi _{i-1}tj_{x,i-1}+\chi _itj_{x,i+1}\right) & -\sin \left(
%\chi _{i-1}tj_{x,i-1}+\chi _itj_{x,i+1}\right) \\ 
%\sin \left( \chi _{i-1}tj_{x,i-1}+\chi _itj_{x,i+1}\right) & \cos \left(
%\chi _{i-1}tj_{x,i-1}+\chi _itj_{x,i+1}\right)
%\end{array}
%\right) 
%%TCIMACRO{\binom{j_{y,i}}{j_{z,i}} }
%BeginExpansion
%{j_{y,i} \choose j_{z,i}}%
%EndExpansion
\label{eq:jt}
\end{equation}
%XXXXXXXXXXXXXX MAYBE LATER XXXXXXXXXXXXXXXXXXXXXXXXXXXXXXXXXXXXXXXXX
%
%At a special time $t=2\pi /\chi ,$ the evolution operator reduces to
%\begin{equation}
%U(2\pi /\chi )=(-i)^N\left[ \sin \left( \frac{\pi \chi ^{\prime }}{2\chi }%
%\right) +i\cos \left( \frac{\pi \chi ^{\prime }}{2\chi }\right) \sigma
%_{x,1}\sigma _{x,N}\right] .
%\end{equation}
%
%For PBC and ABC ($\chi ^{\prime }=\pm \chi $), the evolution operators
%become constants $\pm (-i)^N.$ That is to say, all the physical properties
%are perodic in time with periodicity $2\pi /\chi .$ However for the OBC, we
%can not draw this conclusion since the evolution operator depends on the
%operators $\sigma _{x,1}\sigma _{x,N}$ and not a constant.
%
%XXXXXXXXXXXXXX MAYBE LATER XXXXXXXXXXXXXXXXXXXXXXXXXXXXXXXXXXXXXXXXX
We assume an initial state where all spins are pointing up, i.e., the $j_{z,i}=\frac
12$ 
eigenstate for all $i$, and Eq. (\ref{eq:jt}) then 
provides the mean values of the collective spin components at later
times:

\begin{eqnarray}
\langle J_x\rangle &=& \langle J_y\rangle = 0 \nonumber \\
\langle J_z\rangle &=&\frac 12\sum_{i=1}^N\cos \left( \frac{\chi _it}2\right)
\cos \left( \frac{\chi _{i+1}t}2\right) .  \label{eq:jzz}
\end{eqnarray}

Since the collective operator $J_x$ commutes with $H,$ it follows that
$\langle J_x^2\rangle$ retains its original value at $t=0$:
$\langle J_x^2\rangle =\frac N4.$ In order to calculate $\langle
J_y^2\rangle ,$ we need to know the the expectation values $\langle
j_{y,i}(t)j_{y,i+k}(t)\rangle$. We always have
$\langle j_{y,i}(t)^2\rangle = \frac 14$,  and by direct calculation
we see that $\langle j_{y,i}(t)j_{y,i+1}(t)\rangle =0.$, 
whereas for $k=2$ we obtain
\begin{eqnarray}
\langle j_{y,i}(t)j_{y,i+2}(t)\rangle =\frac 14 \cos \left( \frac{\chi
    _{i-1}t 
}2\right) \sin \left( \frac{\chi _it}2\right) 
 \sin \left( \frac{\chi _{i+1}t}
2\right) \cos \left( \frac{\chi _{i+2}t}2\right) .
\end{eqnarray} 
From Eq.~(\ref{eq:jt}) we see that $\langle
j_{y,i}(t)j_{y,i+k}(t)$  vanishes for $k\geq 3$, and we thus obtain
\begin{eqnarray}
\langle J_y^2\rangle =\frac N4+\frac 12\sum_{i=1}^N\cos \left( \frac{\chi
    _{i-1}t 
}2\right) \sin \left( \frac{\chi _{i}t}2\right)
 \sin \left( \frac{\chi 
_{i+1}t}2\right) \cos \left( \frac{\chi _{i+2}t}2\right),  
\label{eq:jyy} 
\end{eqnarray}
and similarly we can determine the expectation value 
\begin{equation}
\langle J_yJ_x+J_xJ_y\rangle =-\frac 12\sum_{i=1}^N\sin \left( \frac{\chi
_i+\chi _{i+1}}2t\right).   \label{eq:jxy2}
\end{equation}

The degree of spin squeezing can be  characterized  by the squeezing
parameter $\xi^2$ for a particular direction of the noise reduction. 
With the mean spin in the $z$ direction, and considering the reduction of
the noise in the component $J_\theta$ the squeezing parameter can be
calculated by
\begin{equation}
\xi_\theta^2=\frac{\cos^2 (\theta)\langle J_x^2\rangle + \sin^2
(\theta)\langle J_y^2\rangle+\sin(\theta)\cos(\theta)\langle J_x J_y +J_y
J_x\rangle}{\langle J_z \rangle^2}, 
\end{equation}
and for the particular component
$J_{\theta=\pi/4}$  the resulting squeezing factor is
\begin{equation}
\xi_{\pi/4} ^2=\frac{N^2+N\sum_{i=1}^N\cos \left( \frac{\chi _it}2\right) \sin
\left( \frac{\chi _{i+1}t}2\right) \sin \left( \frac{\chi _{i+2}t}2\right)
\cos \left( \frac{\chi _{i+3}t}2\right) -N\sum_{i=1}^N\sin \left( \frac{\chi
_i+\chi _{i+1}}2t\right) }{\left[ \sum_{i=1}^N\cos \left( \frac{\chi _it}2%
\right) \cos \left( \frac{\chi _{i+1}t}2\right) \right] ^2},
\label{eq:xigen}
\end{equation}
which is a general expression for any set of values for the coupling 
coefficients $\chi _i.$ 
For a uniform closed chain (all $\chi_i$ identical), 
the above equation reduces to
\begin{eqnarray}
\xi_{\pi/4}^2=\frac{1+0.25 \sin^2(\chi t)-\sin(\chi t)}{\cos^4(\chi t/2)}
\label{squnif}
\end{eqnarray}
as found in \cite{Soerensen1}.

\subsection{Ising model with few spins}

In the above expressions, coefficients $\chi_{i+k}$ with $i+k > N$ assume in
a cyclic manner the values of the coupling among the first spins in the chain.
It should accordingly be noted that Eq.(\ref{eq:xigen}) is not
valid if $N\leq 4$ , for which special expressions apply.

In the Ising model with only two spins $H_2=2\hbar \chi j_{x,1}j_{x,2}$, the
related expectation values are obtained as $\langle J_x\rangle =\langle
J_y\rangle =0,$ $\langle J_z\rangle =\cos \left( \chi t\right) ,\langle
J_x^2\rangle =\langle J_y^2\rangle =1/2,$ and $\langle J_xJ_y+J_yJ_x\rangle
=-\sin \left( \chi t\right) ,$ from which we obtain the squeezing parameter
\begin{equation}
\xi_{\pi/4} ^2=\frac{1-\sin \left( \chi t\right) }{\cos ^2\left( \chi
    t\right) } .
\label{eq:xi2}
\end{equation}
The system can be squeezed, and the maximum squeezing $\xi_{\pi/4} ^2=0.5$ occurs
when $t=\pi /\left( 2\chi \right)$.

For the Ising model with three spins $H_3=\hbar \sum_{i=1}^3\chi _ij_{x,i}j_{x,i+1},$
the squeezing parameter is
\begin{equation}
\xi_{\pi/4} ^2=\frac{9+3\sum_{i=1}^3\sin \left( \frac{\chi _it}2\right)
  \sin \left(  
\frac{\chi _{i+1}t}2\right) -3\sum_{i=1}^3\sin \left( \frac{\chi _i+\chi
_{i+1}}2t\right) }{\left[ \sum_{i=1}^3\cos \left( \frac{\chi _it}2\right)
\cos \left( \frac{\chi _{i+1}t}2\right) \right] ^2},
\end{equation}
and for a uniform closed chain 
this equation reduces to
\begin{equation}
\xi_{\pi/4} ^2=\frac{1+\sin ^2\left( \frac{\chi t}2\right) -\sin \left( \chi
t\right) }{\cos ^4\left( \frac{\chi t}2\right) },  \label{eq:xi33}
\end{equation}
which is different from Eq.(\ref{squnif}).

Finally, in the Ising model with four spins 
$H_4=\hbar \sum_{i=1}^4\chi _ij_{x,i}j_{x,i+1}.$ The
squeezing parameter is given by
\begin{equation}
\xi_{\pi/4} ^2=\frac{8(\langle J_x^2\rangle +\langle J_y^2\rangle
)-4\sum_{i=1}^4\sin \left( \frac{\chi _{i}+\chi _{i+1}}2t\right) }{\left[
\sum_{i=1}^4\cos \left( \frac{\chi_{i}t}2\right) \cos \left( \frac{\chi
_{i+1}t}2\right) \right] ^2},
\label{eq:xi4}
\end{equation}
where
\begin{eqnarray}
\langle J_x^2\rangle +\langle J_y^2\rangle
=2+\frac{1}{2} \sum_{i=1}^4 \sin{\left(\frac{\chi_i t}{2}\right)}
\sin{\left(\frac{\chi_{i+1} t}{2}\right)} \cos{\left(\frac{\chi_{i+2}
      t}{2}\right)} \cos{\left(\frac{\chi_{i+3} t}{2}\right)}.
%  &=&2+\frac 12[\sin \left( \frac{%
% \chi _1t}2\right) \sin \left( \frac{\chi _2t}2\right) \cos \left( \frac{\chi
% _3t}2\right) \cos \left( \frac{\chi _4t}2\right)   \nonumber \\
% &&+\cos \left( \frac{\chi _1t}2\right) \cos \left( \frac{\chi _2t}2\right)
% \sin \left( \frac{\chi _3t}2\right) \sin \left( \frac{\chi _4t}2\right)  
% \nonumber \\
% &&+\sin \left( \frac{\chi _1t}2\right) \sin \left( \frac{\chi _4t}2\right)
% \cos \left( \frac{\chi _2t}2\right) \cos \left( \frac{\chi _3t}2\right)  
% \nonumber \\
% &&+\cos \left( \frac{\chi _1t}2\right) \cos \left( \frac{\chi _4t}2\right)
% \sin \left( \frac{\chi _2t}2\right) \sin \left( \frac{\chi _3t}2\right) ].
\end{eqnarray}
For a uniform chain 
Eq. (\ref{eq:xi4}) actually reduces to Eq.(\ref{squnif}),
but for general coupling constants $%
\chi _i,$ Eq. (\ref{eq:xigen}) is only valid for $N\geq 5.$

Fig. 1 is a plot of the squeezing parameter as a function of time for few
spins in a uniform chain.  
It shows that maximum squeezing occurs for $N=2$ 
and that the degree and the temporal range of squeezing are both small 
for $N=3$ in comparison with the other cases.

\begin{figure}
\begin{center}
 \epsfig{file=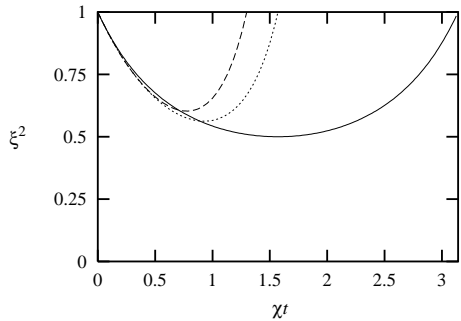,angle=270,width=6cm}
\end{center}
\vspace{0.5cm}
 \caption{Squeezing parameter as function of time in the Ising model
with uniform coupling coefficients. 
The solid curve is for $N=2$ spins, 
the dashed cuve is for $N=3$, and the dotted curve is for 
$N=4$ spins or more.}
\end{figure}

\subsection{Dimerized and random chains} 

It is interesting to study a dimerized Ising chain with even number 
of spins $N=2M$, where the coupling constants are chosen as $\chi
_i=\chi [1+(-1)^{i+1}\delta ]$. We
have two values of the couplings $\chi _o=\chi \left( 1+\delta \right) $ for
odd $i$ and $\chi _e=\chi \left( 1-\delta \right) $ for even $i.$
Such systems appear naturally in heterogeneous structures
where  every second site is occupied with one type of 
qubit/particle, and their mutual communication is provided through
intermediate particles acting as short range data-bus elements in
e.g. the quantum computer. Two overlapping optical lattices with two
different kinds of atoms or atoms in two different ground states
may be moved relative to each other in order to establish
the dimerized chain  Hamiltonian.

For the dimerized Ising chain, Eq.(\ref{eq:xigen}) reduces to
\begin{equation}
\xi_{\pi/4} ^2=\frac{1+0.25\sin \left( \chi _ot\right) \sin \left( \chi
    _et\right) 
-\sin \left( \chi t\right) }{\cos ^2\left( \frac{\chi _ot}2\right) \cos
^2\left( \frac{\chi _et}2\right) },  \label{eq:xidem}
\end{equation}
and we see that the spin squeezing does not depend on the number of
atoms. Obviously Eq.(\ref{eq:xidem}) reduces to 
(\ref{squnif}) in the limit $\delta \rightarrow 0$, and in the limit
$\delta\rightarrow 1$ it gives Eq.~(\ref{eq:xi2}). 
Fig. 2 is a plot of the
squeezing parameter as a function of the parameter $\delta$. We see that both
the squeezing range and degree of squeezing increases 
as $\delta$ approaches unity, i.e., the introduction of a dimerized coupling
makes the squeezing better. For $\delta$ larger than unity, the
denominator vanishes within the time interval displayed in the figure, 
as shown for 
$\delta=1.1$ which causes a divergence of the squeezing parameter at
$\chi t=\pi/2.1 = 1.496$.

\begin{figure}
\begin{center}
 \epsfig{file=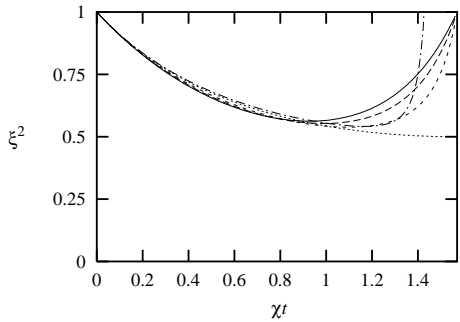,angle=270,width=6cm}
\end{center}
\vspace{0.5cm}
 \caption{Squeezing parameter as function of time in the Ising model
with dimerized coupling coefficients. The curves are obtained with
the following values of 
$\delta$: 0 (solid curve) , 0.5 (dashed), 0.75 (short-dashed), 
1 (dotted), and 1.1 (dot-dashed).
The curves
with $\delta=0$ and $\delta=1$ are recovered in Fig. 1.
	}
\end{figure}

Let us finally study a random chain model in which  spins interact with 
their neighbours with a fixed coupling constant $\chi$, but only with
probability $p,$ i.e., any coupling constant is $\chi $ with probability $p$
and zero with probability $1-p.$  This model corresponds, e.g., to 
the atomic lattice system with a non-unit filling fraction, so that
empty lattice sites appear. 
Different assumptions may be made about the correlations between
different lattice sites, but let us for simplicity consider the 
case where the occupancies are completely uncorrelated.
Simulations for such a model were
presented in Ref.\cite{Soerensen1}, but  we can
now present analytical results (for simplicity presented only for the
$\theta=\pi/4$ direction).

We introduce $\mu = \chi t/2$, in terms of which we can write the
mean values of trigonometric functions,
\begin{eqnarray}
\overline{\cos(\frac{\chi_i t}{2})} &=& p \cos\mu+(1-p)\nonumber \\
\overline{\sin(\frac{\chi_i t}{2})} &=& p \sin\mu \nonumber \\
\overline{\sin(\frac{\chi_i+\chi_{i+1}}{2}t)} &=& p^2
\sin2\mu+2p(1-p)\sin\mu.
\label{trig}
\end{eqnarray}

We assume a very large number of particles, so that the mean value of
Eq.(\ref{eq:xigen}) over different realizations of the spin lattice, is
effectively 
obtained by considering only a single realization consisting of a
very large number of particles. The enumerator is obtained 
simply as products of terms like in (\ref{trig}).
The denominator of Eq.(16) involves the mean value of the square of a 
sum of products of $\cos$-functions, and in the limit of large $N$, 
it is dominated by terms which are products of four uncorrelated
$\cos$-functions, and we obtain
\begin{eqnarray}
\xi_{\pi/4}^2 = \frac{1+\left(1-p(1-\cos\mu)\right)^2p^2\sin^2\mu-
\left(p^2\sin2\mu+2p(1-p)\sin\mu\right)}{\left(1-p(1-\cos\mu)\right)^4}
\end{eqnarray}
As shown in Fig. 3 and observed already in
\cite{Soerensen1}, spin squeezing is obtained also in the case
of a random filling.

\begin{figure}
\begin{center}
 \epsfig{file=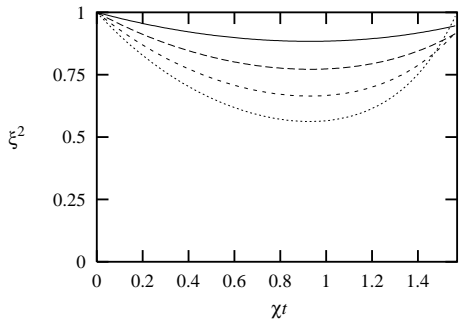,angle=270,width=6cm}
\end{center}
\vspace{0.5cm}
 \caption{Squeezing parameter as function of time in the Ising model
with random coupling coefficients. Every coupling coeffecient $\chi_i$
is a stochastic variable, attaining the value $\chi$ with probability
$p$ and the value $0$ with probability $(1-p)$.
The curves are obtained with
the following values of $p$: 0.25 (solid curve), 0.5 (dashed), 0.75
(short-dashed), and  1 (dotted). 
	}
\end{figure}

\section{Conclusion}

We have obtained general expressions for squeezing 
in different models of interacting spins. In the general model
considered in section \ref{general} we showed that almost any kind of
spin-spin interaction can be used to create squeezed states. As long as the
coupling matrix in Eq.~(\ref{m}) is not proportional to the identity, spin
squeezing can be produced.
In the 
one-dimensional Ising model with arbitrary nearest neighbour coupling 
constants we obtained analytical results for all times, and 
we addressed  the case of constant couplings and 
different kinds of departure from constant couplings of the spins.
We found that the amount of spin squeezing was larger in
a dimerized model with periodically varying coupling coefficients than
in a homogeneous model. 

Among many possible generalizations of the
present work we wish to mention  long-range interactions and 
interactions in higher dimensions, and, {\it e.g.}, the question
whether extensions of the dimerized model to these
cases are also superiour over homogeneous couplings.

As mentioned in the introduction, spin squeezing is both a useful
property in itself \cite{Wineland} and a signature of entanglement
\cite{Soerensen3,Soerensen2}.  It can be 
accomplished and detected in different physical systems without
the need for experimental access to individual spins, and as such it
can be used to test, {\it e.g.}, spintronics proposals for quantum information
processing.


\begin{thebibliography}{99}

\bibitem{Squeezed}  D. F. Walls and G. J. Milburn, {\it Quantum Optics}
(Springer, Berlin 1994).

\bibitem{Kitagawa}  M. Kitagawa and M. Ueda, \pra {\bf 47}, 5138 (1993).

\bibitem{Wineland}  D. J. Wineland {\it et al}., \pra {\bf 46}, 11 (1992); 
\pra {\bf 46}, R6797 (1992); \pra {\bf 50}, 67 (1994).

\bibitem{Agarwal}  G. S. Agarwal and R. R. Puri, Phys. Rev. A {\bf 41}, 3782
(1990); Phys. Rev. A {\bf 49}, 4968 (1994).

\bibitem{Kuzmich1}  A. Kuzmich, K. M\o lmer, and E. S. Polzik, Phys. Rev.
Lett. {\bf 79}, 4782 (1997); A. Kozhekin, K. M\o lmer, and E. S. Polzik,
Phys. Rev. A {\bf 62}, 033809 (2000).


\bibitem{Lukin}  M. D. Lukin, S. F. Yelin, and M. Fleischhauer, Phys. Rev. 
Lett. {\bf \ 84}, 4232 (2000).

\bibitem{Vernac}  L. Vernac, M. Pinard, and E. Giacobino, Phys. Rev. A {\bf 
62}, 063812 (2000).

\bibitem{Kuzmichqnd}  A. Kuzmich, N. P. Bigelow, and L. Mandel, Europhys. 
Lett. {\bf 43}, 481 (1998).

\bibitem{Soerensen1}  A. S\o rensen and K. M\o lmer, Phys. Rev. Lett. {\bf
    83},  2274 (1999).

\bibitem{Soerensen3}  A. S\o rensen, L.-M. Duan, J. I. Cirac, and P. Zoller,
Nature {\bf 409}, 63 (2001); L. -M. Duan, A. S\o rensen, J. I. Cirac, and
P. Zoller, Phys. Rev. Lett. {\bf 85}, 3991 (2000).

\bibitem{Soerensen2}  A. S. S\o rensen and K. M\o lmer, \prl {\bf 86}, 4431
  (2001). 

\bibitem{hald}  J. Hald, J. L. S\o rensen, C. Schori, and E. S. Polzik, Phys.
Rev. Lett. {\bf 83}, 1319 (1999).

\bibitem{Kuzmichexp} A. Kuzmich, L. Mandel, and N. P. Bigelow, Phys. Rev.
Lett. {\bf 85}, 1594 (2000).

\bibitem{kasevich} C. Orzel, A. K. Tuchman, M. L. Fenselau, M. Yasuda,
   and M. A. Kasevich, Science {\bf 291}, 2386 (2001).

\bibitem{risq} K. M\o lmer and A. S\o rensen, J. Mod. Opt. {\bf 47}, 
  2515 (2000). 
 
\bibitem{innsbruck_lattice} D.~Jaksch, H.-J.~Briegel, J.~I.~Cirac,
  C.~W.~Gardiner, 
  and P.~Zoller, Phys. Rev. Lett. {\bf 82}, 1975
  (1999).

\bibitem{jessen} G.~K.~Brennen , C.~M.~Caves, P.~S.~Jessen, and
  I.~H.~Deutsch, Phys. Rev. Lett. {\bf 82},
  1060 (1999). 

\bibitem{briegel} R. Raussendorf and H.-J. Briegel, \prl {\bf 86},
 5188 (2001); 
 H.-J. Briegel and R. Raussendorf, Phys. Rev. Lett. {\bf 86}, 910
 (2001). 


%\bibitem{Trifonov}  D. A. Trifonov

\end{thebibliography}
\end{document}